\documentstyle[11pt, aaspp4]{article}

\newcommand{\ads}{accretion disks } 
\begin{document}

\title{On the relative surface density change of thermally unstable \ads }

\author{Xue-Bing Wu\altaffilmark{1,2}}

\altaffiltext{1}{Institute of Theoretical Physics, Chinese Academy of 
Sciences, Beijing
 100080, China; email: wuxb@itp.ac.cn}
\altaffiltext{2}{Beijing Astronomical Observatory, Chinese Academy of Sciences,
 Beijing
 100080, China} 

\begin{abstract}

The relations among the relative changes of surface density, temperature, 
disk height and vertical integrated pressure in three kinds of thermally 
unstable \ads  were quantitatively
investigated by assuming local perturbations. The surface density change was 
found to be very small in the
long perturbation wavelength case but can not be ignored in the short 
wavelength
case. It becomes significant in
an optically thin, radiative cooling dominated disk when the perturbation 
wavelength is shorter than 15H (H is the scale height of disk) and in a 
geometrically thin,
optically thick and radiation pressure dominated disk when the perturbation 
wavelength is shorter than 50H. In an optically thick, advection-dominated 
disk,
 which is thermally unstable against short wavelength perturbations,
the relative surface density change is much larger. We proved the positive 
correlation between the changes of surface density and temperature in an 
optically thick, advection-dominated disk which was previously claimed to be
the essential point of its thermal instability. Moreover, we 
found an anticorrelation between the changes of disk height and temperature
in an optically thick, advection-dominated disk. This is the natural result
of the absence of appreciable vertical integrated pressure change.
\end{abstract}
\keywords {accretion, \ads --- black hole physics --- instabilities}

\section{INTRODUCTION}

Shortly after the construction of the geometrically thin, optically thick 
 accretion disk model in early 1970s (Shakura \& Sunyaev 1973; Novikov \&
Throne 1973), the stability analyses indicated that the inner part of such a 
disk,
where radiation dominates the pressure and  electron
 scattering dominates the opacity, is thermally and secularly unstable (Shakura \& 
 Sunyaev 1976;
Lightman \& Eardley 1974). The optically thin disk model, proposed at first to
 account
 for the hard X-ray radiation of Cyg X-1 by Shapiro, Lightman \& Eardley 
 (1976),
was also shown to be thermally unstable (Pringle, Rees \& Pacholczyk 1976;
Pringle 1976). The general criteria of the stability of \ads were derived by
Piran (1978) who considered various kinds of cooling processes in the disks.
In last two decades, the thermal instability of accretion disks has been 
extensively studied with important applications to many time varying phenomena
 in dwarf novae, X-ray binaries and even AGNs, though its rapid growing also
may lead to the breakdown of the disk equilibrium.

The early works we mentioned above all assumed that the surface density of the
 disk does not change when the thermal instability was addressed. This is a
good approximation when the geometrically thin disk and the long wavelength
perturbations were considered. In this case the effect of pressure force is 
minor
and the radial balance is realized by 
the balance between the centrifugal and gravitational forces. A temperature 
change associated with the thermal mode occurs with no motion in the
horizontal plane. The gas expands or shrinks only in the vertical direction 
so
that the vertical hydrostatic balance is realized without changing the surface
density. This is the essential point of the thermal instability in a
 geometrically
thin disk (Kato, Abramowicz \& Chen 1996, hereafter KAC). However, a hot 
optically thin disk
may be not geometrically thin (Shapiro et al. 1976), the previous stability
analyses assuming no surface density change are in doubt in such a disk. 
Moreover,
an optically thick, advection-dominated disk, which is usually geometrically 
slim, was recently shown to be thermally unstable against short wavelength 
perturbations
(KAC; Wu \& Li 1996). Evidently most of the previous analyses, usually dealing 
with 
the long wavelength perturbations, are not suitable in this case.
KAC also pointed out that in an advection-dominated disk the thermal 
instability
occurs with large surface density change but with only small pressure change. 
This
is quite different from the case of a geometrically thin disk where the thermal
instability occurs with no appreciable surface density change. In addition,
if the surface density change is strongly associated with the temperature
change, some previous stability analyses, such as that simply assuming the 
surface density unchanged and comparing the cooling and heating rate near the 
equilibrium curve (usually the $\dot{M}(\Sigma)$ curve), would be not 
self-consistent (Chen et al., 1995). Therefore, a clear understanding of the 
relation between
the surface density change and the temperature change is important.

In this paper, we will quantitatively investigate the relations among the 
changes of surface density, temperature, disk height and vertical integrated
pressure in three kinds of thermally unstable accretion disks. Their 
dependences on the perturbation wavelength will be also
calculated. Our present study serves as an extension of a recent work by Wu \& 
Li (1996) (hereafter Paper I), where the linear stability properties of 
various kinds of 
disks were discussed
by numerically solving a general dispersion relation. Our quantitative analyses
will prove
some qualitative results obtained previously and also reach some new 
conclusions.

\section{BASIC EQUATIONS}

The general time-dependent hydrodynamic equations which are usually involved
to describe the viscous accretion models have been given by many authors. The
vertical integrated equations in a cylindrical system of coordinates ($r,
\varphi, z$) 
were also summerized in the Eqs. (1)-(4) of Paper I where the disk was assumed 
to be 
axisymmetric 
and non-self-gravitating.  
The radial perturbations were assumed of the
 form $(
\delta V_{r}, \delta \Omega, \delta \Sigma, \delta T)\sim e^{i(\omega t-kr)}$,
where $V_{r}, \Omega, \Sigma$ and $T$ are the radial velocity, angular 
velocity, surface 
density and temperature. $k$ is the perturbation wavenumber defined by 
$k=2\pi/\lambda$, 
$\lambda$ is
the perturbation wavelength. 

In this work we consider the local approximation, which means
$\lambda < r$. The validity
of the vertically integrated equations requires $kV_{r} < \Omega_{k}$, where
$\Omega_{k}$ is the Keplerian angular velocity. Since
$V_{r} \sim \alpha {c_{s}}^{2}/r\Omega_{k}$ where $\alpha$ is the viscosity 
parameter and $c_s$ is the local sound speed, the requirements above can be 
written as
$\frac{r}{H}> \frac{\lambda}{H} > 2\pi \alpha \frac{H}{r}$.
We can see clearly that this inequality is well satisfied for a geometrically 
thin
accretion disk even if we set $\lambda/H$ and $\alpha$ in a wide range, such 
as $\lambda/H$ from 1 to
100 and $\alpha$ from 0.001 to 1. However, for a geometrically slim disk, 
where $H/r \le 1$, it can be satisfied
only when $\alpha$ is sufficiently small. The range of $\lambda/H$ also moves 
to
the smaller value, such as from 0.01 to 2 if $\alpha$ is about 0.001 and $H/r$
is about 0.6. In addition, the validity of vertical integrated equations also 
requires the growth rates of
unstable modes are less than the angular velocity (see also KAC). In order 
to
get reasonable and self-consistent results, we present
our discussions in this paper following all these restrictions.

If we consider the local approximation, the linearized perturbed equations 
can be written as:
\begin{equation}
{\bf A}\frac{\delta \Sigma}{\Sigma}+{\bf B}\frac{\delta V_r}{\Omega_k r}
+{\bf C}\frac{\delta \Omega}{\Omega_k} +{\bf D}\frac{\delta T}{T}=0
\end{equation}
where {\bf A, B, C} and {\bf D} are four-component matrices and are functions 
of
$\tilde \sigma$ ($\tilde \sigma=\sigma/\Omega_k$ where 
$\sigma=i(\omega-kV_{r})$)
and other disk parameters (see Paper I for details). By setting the 
determinant of 
the coefficients in above perturbed equation 
to zero, we can get a dispersion relation:
\begin{equation}
a_{1}{\tilde{\sigma}}^{4}+a_{2}{\tilde{\sigma}}^{3}+a_{3}{\tilde{\sigma}}^{2}+
a_{4}\tilde{\sigma}+a_{5}=0,
\end{equation}
where $a_{i} (i=1,...,5)$ is the coefficients given by the functions of disk
parameters and perturbation wavelength. The four kinds of solutions of the 
dispersion 
relation are related with the stability properties of two inertial-acoustic 
modes, thermal 
and viscous modes in the disk.
The real parts of the solutions correspond to the growth rates of the 
perturbation modes and the imaginary parts correspond to their propagating 
properties. If the real part corresponding to certain mode is positive, this 
mode will 
be unstable. Paper I has numerically solved the dispersion
relation and obtained the stability properties of accretion disks with 
different
disk structures. It also proved that there are three kinds of thermally
unstable accretion disks, namely, the optically thin, radiative cooling 
dominated disk, 
the optically thick, radiative cooling and radiation pressure
dominated disk and the optically thick, advection-dominated disk.

The relation between the relative changes of surface density and temperature
in thermally unstable disks can be calculated by solving Eq. (1) if some 
terms such as $\tilde\sigma$, $\lambda$ and other disk parameters are 
specified.
In this paper, we will obtain this relation based on the solutions given by 
Paper I to 
the thermal modes of three kinds of thermally unstable accretion disks.
 
Before solving the Eq. (1), we first mention some important relations
among the changes of surface density and other disk parameters in
accretion disks. For convenience we define some quantities as follows:
\begin{equation}
\delta_{\Sigma T}=\frac{\delta \Sigma}{\Sigma}/\frac{\delta T}{T},~~~
\delta_{\Sigma W}=\frac{\delta \Sigma}{\Sigma}/\frac{\delta W}{W},~~~
\delta_{\Sigma H}=\frac{\delta \Sigma}{\Sigma}/\frac{\delta H}{H}
\end{equation}
where W is the vertical integrated pressure. 
Above quantities 
describe the ratios of the relative surface density change and the relative 
changes of 
temperature, vertical integrated pressure and disk height respectively.

In an optically thin, radiative cooling dominated disk, we have $H=c_s/
\Omega_k\propto 
T^{1/2}$, $\rho=\Sigma/2H \propto\Sigma T^{-1/2}$ and $W=2pH\propto \Sigma T$.
The local sound speed is given by $c_s=p/\rho$ and p, $\rho$ are 
the
pressure and density respectively.
Thus, the variation of W and H can be approximated by $\frac{\delta H}{H}=
\frac{1}{2}\frac{\delta T}{T}$ and $\frac{\delta W}{W}=\frac{\delta \Sigma}
{\Sigma}+\frac{\delta T}{T}$. That is,
\begin{equation}
  \delta_{\Sigma H}=2\delta_{\Sigma T},~~~\delta_{\Sigma W}=\frac{1}{1+1/
  \delta_{\Sigma T}}.
\end{equation}
 In an optically thick, radiation pressure dominated disk, we can write 
 $H\propto 
\Sigma^{-1}T^4$ and $W\propto \Sigma^{-1} T^8$. Thus, we have:
\begin{equation}
  \delta_{\Sigma H}=\frac{1}{4/\delta_{\Sigma T}-1},~~~\delta_{\Sigma W}=\frac
  {1}{8/
  \delta_{\Sigma T}-1}.
\end{equation}
These relations exist not only in a radiative cooling dominated disk but also
in an advection-dominated disk. This is because only the equations of 
hydrostatic equilibrium and mass conservation and the equation of state were 
used to
derive above relations, and these equations are the same for the two kinds of 
optically thick
disks. The energy transport equation, which is different in these two kinds of optically thick disks, does not enter into above relations.
Therefore from Eqs. (4) and (5) we can easily obtain the relative 
changes of 
surface density to disk height and
vertical investigated pressure as long as $\delta_{\Sigma T}$ is given. 

\section {NUMERICAL RESULTS}

Following we will
solve Eq. (1) using the solution corresponding to thermal unstable mode 
obtained
in Paper I, and derive the relative changes of surface density to temperature,
disk height and vertical integrated pressure in three kinds of unstable 
accretion disks. 

Fig. 1 shows the case of a hot optically thin, radiative cooling dominated 
disk.
For completeness in (a) we show the stability properties of a disk where the 
cooling 
mechanism is bremsstrahlung (It is similar as Fig. 1 in Paper I but with a 
minor 
numerical correction).  
We can clearly see that the thermal mode is always unstable and the viscous 
mode is 
always stable. The inertial-acoustic modes are slightly unstable to long 
wavelength 
perturbations but stable to short
wavelength perturbations. If the disk is a hot two temperature one and the 
Compton 
scattering dominates radiative cooling, the stability properties are quite 
similar 
with those in a
bremsstrahlung disk (Pringle 1976). In (b) we show the the relative changes of
surface density to temperature, disk height and vertical integrated pressure 
corresponding 
to the unstable thermal mode.
Evidently in long perturbation wavelength case, the surface density nearly 
does
not change. But the surface density change becomes significant if the 
perturbation wavelength 
is shorter than 15H. This result shows that some previous stability analyses 
of hot optically 
thin disks where the surface density was assumed to be constant, such as 
Pringle, Rees
\& Pacholczyk (1976), Pringle (1976) and Piran (1978) are correct only in the
long perturbation wavelength case. If the hot optically thin disk is not 
geometrically thin, 
the linear local stability analysis is valid only when
the viscosity coefficient $\alpha$ is small and the perturbation wavelength is 
short (see 
KAC and Paper I). Thus, some previous analyses we mentioned above are
not appropriate in this case where the surface density change is significant 
with the 
temperature perturbations. Fig. 1(b) also shows that  $\delta_{\Sigma T}$, 
$\delta_{\Sigma H}$ 
and $\delta_{\Sigma W}$ are all negative. It means 
that with positive temperature perturbations the surface density decreases
 slightly
but the disk height and vertical integrated pressure increase substantially.
Actually in the long perturbation wavelength limit, we have $\frac{\delta 
\Sigma}
{\Sigma}\sim 0$, $\frac{\delta H}{H}=
\frac{1}{2}\frac{\delta T}{T}$ and $\frac{\delta W}{W}\sim \frac{\delta T}{T}$.
In the short perturbation wavelength limit, we note that Fig. 1(b) remains 
valid even when $\alpha$ is very small and $H/r\le1$, and we have  
$\frac{\delta \Sigma}
{\Sigma}\sim -\frac{\delta T}{T}$, $\frac{\delta H}{H}=
\frac{1}{2}\frac{\delta T}{T}$ and $\frac{\delta W}{W}\sim 0$. We see that in 
a hot
optically thin disk, the thermal instability in the long perturbation 
wavelength limit 
is different from that in the short perturbation wavelength limit. In the
former case, the thermal instability occurs with no appreciable surface density
change. In the later case, however, the thermal instability occurs with no
appreciable vertical integrated pressure change. 

Fig. 2 shows the case of an optically thick, radiative cooling and radiation 
pressure 
dominated disk, which is usually geometrically thin. From (a) we see that the 
disk is thermally and viscously
unstable in long perturbation wavelength case but is acoustically unstable in the
short perturbation wavelength case (see also Fig. 9 in Paper I). The 
non-traveling, 
unstable thermal mode exists when $\lambda/H>32$ if $\alpha$ is taken as 0.01. 
In (b) we see
that the relative change of surface density is small comparing with the 
relative
changes of disk height and vertical 
integrated pressure,  but it is significant comparing with the relative 
 temperature change if the 
perturbation wavelength is shorter than 50H.
Evidently the assumption that the surface density is constant is generally not 
appropriate 
in this case except in the long perturbation wavelength limit. When the 
perturbation wavelength 
is longer than 32H,
the unstable thermal mode is not a traveling mode and we see that 
$\delta_{\Sigma T}$, 
$\delta_{\Sigma H}$ and $\delta_{\Sigma W}$  are all negative. It means that 
with a 
temperature increasement the surface density decreases but the disk height 
and vertical 
integrated pressure increase, which is quite similar with the result in a hot 
optically 
thin disk. However,
the increases of disk height and vertical integrated pressure in an optically
thick, radiative cooling dominated disk are much larger than those in a hot
optically thin disk if we assume a temperature increase. For example, in the 
long perturbation 
wavelength limit where $\frac{\delta \Sigma}
{\Sigma}\sim 0$, we have $\frac{\delta H}{H}\sim
4\frac{\delta T}{T}$  and $\frac{\delta W}{W}\sim 8\frac{\delta T}{T}$. When
the perturbation wavelength is shorter than 32H, the unstable thermal mode
mixes with the viscous mode and becomes complex. It is a traveling mode which
is different from the ordinary thermal mode. We will not discuss it in detail 
in this
paper. 

Fig. 3 shows the case of an optically thick, radiation pressure and advection
dominated disk, which is usually geometrically slim. Due to the local
restriction, we will discuss the thermal instability against short wavelength
perturbations and assume $\alpha$ is very small. From (a) we see that the 
thermal mode is always unstable and
the viscous mode is stable (see also Fig. 12 in paper I). The disk is more 
thermally unstable 
if the thermal diffusion effect is considered. In (b) we
clearly see that the surface density change is very important in this case.
We have $\delta_{\Sigma T}\sim 8$, $\delta_{\Sigma H}\sim -2$ and $\delta_
{\Sigma W}\sim 
+\infty$. Although the inclusion of thermal diffusion
has a significant effect to enhance the thermal instability, it does not
affect the surface density change very much. With a positive temperature 
increase associated 
with the thermal perturbation, the surface density increases significantly but 
the disk height
decreases and the vertical integrated pressure nearly does not change. 
Here we also have $\frac{\delta H}{H}\sim
-4\frac{\delta T}{T}$  and $\frac{\delta W}{W}\sim 0$. This is
evidently quite different from the results in above two cases. It also implies that
the essential point of thermal instability in an advection-dominated disk is
different from that in a hot optically thin disk or a geometrically thin,
optically thick and radiative cooling dominated disk.

Our calculations indicate that in the long perturbation wavelength limit
 $\mid \frac{\delta \Sigma}{\Sigma}\mid$ is much less than 
$\mid \frac{\delta T}{T}\mid$, $\mid \frac{\delta H}{H}\mid$ and 
$\mid \frac{\delta W}{W}\mid$ either in a hot optically thin disk or in
a geometrically thin, optically thick and radiation pressure dominated disk.
However, in the short perturbation wavelength case, we have  $\mid \frac
{\delta \Sigma}
{\Sigma}\mid < \mid \frac{\delta T}{T}\mid$,  $\mid \frac{\delta \Sigma}
{\Sigma}\mid > 
\mid \frac{\delta H}{H}\mid$ and $\mid \frac{\delta \Sigma}{\Sigma}\mid > 
\mid 
\frac{\delta W}{W}\mid$ in a hot optically thin disk,
and  $\mid \frac{\delta \Sigma}{\Sigma}\mid > \mid \frac{\delta T}{T}\mid$,  
$\mid 
\frac{\delta \Sigma}{\Sigma}\mid << \mid \frac{\delta H}{H}\mid$ and $\mid 
\frac{\delta 
\Sigma}{\Sigma}\mid << \mid \frac{\delta W}{W}\mid$ in a 
geometrically thin, optically thick and radiation pressure dominated disk if
the thermal mode is not a traveling mode. We can see that the behaviors of 
surface density
 change of these two kinds of disks are different in the
short perturbation wavelength case, though they are similar in the long 
perturbation wavelength 
limit.  In an optically thick, advection-dominated disk,
we have $\mid \frac{\delta \Sigma}{\Sigma}\mid >> \mid \frac{\delta T}{T}\mid$,
  $\mid 
\frac{\delta \Sigma}{\Sigma}\mid > \mid \frac{\delta H}{H}\mid$ and $\mid 
\frac{\delta 
\Sigma}{\Sigma}\mid >> \mid \frac{\delta W}{W}\mid$. These are different from 
those in 
either the long perturbation wavelength case or
short perturbation wavelength case of other two kinds of thermally unstable
disks. Our results show that the surface density change can be ignored in 
investigating the thermal instability against long wavelength perturbations 
but
can not be ignored in investigating the thermal instability against short 
wavelength 
perturbations especially in an optically thick, advection-dominated disk.  

Another very interesting result of our investigation is the anticorrelation of
the changes of the disk height and temperature in a thermally unstable 
advection-dominated 
disk. This has not been recognized in the analytic analysis of KAC. In fact, 
it is a natural 
result of the effect that the thermal instability occurs without significant 
vertical 
integrated pressure change in an optically thick, advection-dominated 
accretion disk where we have
 $W\propto T^4H$. 
The cause of thermal instability can be also understood as follows: Fig. 3(b) 
have indicated 
that 
 $\delta_{\Sigma T}\sim 8$.  If we assume an
increase of temperature, it will lead to the much larger increase of surface
density, which will result in the decrease of disk height since  
$H=c_s/\Omega_k
\propto\Sigma^{-1}T^4$. Moreover, in an optically thick advection-dominated 
disk there is 
no appreciable vertical integrated pressure change. The decrease of disk height
 will act  
to amplify the  increasement of temperature since $W\propto T^4H$.  Therefore, 
we can clearly 
see that the 
positive correlation of the changes of surface density and temperature and
the absence of the change of vertical integrated pressure are two essential 
points of thermal 
instability in an advection-dominated
disk. These properties are quite different from those in other
two kinds of thermally unstable disks without advection. 

\section{DISCUSSIONS}

Our
quantitative investigations not only proved that the essential point of 
thermal instability in
 an advection-dominated disk is different from that in a disk without advection
  which has been
  previously indicated by KAC, but also derived
some new conclusions. First, the dependence of surface density change on the
perturbation wavelength is obtained for three kinds of thermally unstable 
accretion disks. We 
showed that the surface density changes are small in the
long perturbation wavelength limit but are not ignorable in the short 
perturbation wavelength 
case. Especially in an optically thick, advection-dominated disk, the surface 
density change 
is much significant against
short wavelength perturbations. Second, the relations among the changes of 
surface density 
and other disk parameters in an advection-dominated disk  are different from 
those in other 
disks without advection. In later
cases, there are anticorrelations among the surface density change and 
the changes of 
temperature, disk height and vertical integrated pressure. In former
case, however, there are anticorrelation between the changes of surface 
density
and disk height but positive correlation among the change of surface density
and the changes of temperature and vertical integrated pressure. Third, we 
found
an anticorrelation between the changes of temperature and disk height
for an optically thick, advection-dominated disk. We have pointed out that it
is a natural effect of the absence of the evident change of the vertical
integrated pressure, which is also the essential point of the thermal 
instability of an 
advection-dominated disk.

Some early stability analyses, where the surface density was often assumed to 
be constant to 
derive the thermal instability criteria, were found to be valid only
in the long wavelength perturbation limit. This is probably the case for a
geometrically thin disk where the gravitational force is balanced by the 
centrifugal force 
and the radial pressure force is small. Thus, the variation 
of temperature only leads to the gas expansion or contraction in the vertical 
direction and 
the thermal instability occurs without appreciable surface density change. 
However, if the 
perturbation wavelength is
not longer enough, the change of surface density must be taken into account 
and
those previous thermal instability criteria are not appropriate. The
change of surface density is the effect of radial expansion or contraction due
to the non-neglectable radial pressure gradient. This usually happens in a 
geometrically slim 
or thick disk if the short wavelength perturbations are
considered. Especially in an optically thick, advection-dominated disk, the 
thermal instability occurs with much more significant surface density change 
than the changes of vertical integrated pressure, temperature and disk height.
We must consider the effects associated with the pressure, velocity and 
advection in the radial direction 
and take into account the surface density change in deriving the thermal 
instability criteria.

Together with the results in KAC and Paper I, our investigations indicate
that the optically thick, advection-dominated disk is thermally unstable
against local short wavelength perturbations. This is quite different from 
some previous implications. For example, Abramowicz et al. (1995) and Chen 
et al. (1995) showed that the upper branch of their S shape curves in 
$\dot{M}-\Sigma$ diagram, which corresponds to the optically thick, 
advection-dominated 
equilibrium, is thermally stable. This result was obtained by comparing the 
heating and cooling rate near the equilibrium curve. As we have mentioned in 
Section 1, however, such kind of analysis usually considered only the 
temperature (or $\dot{M}$) change and assumed that the local surface density 
is constant. This is not
appropriate because the surface density change associated with the temperature
 change is significant in an optically thick, advection dominated disk and can 
not be ignored. Thus, although the upper branch of the S shape curve has a 
positive slope, it only means that the optically thick, advection dominated 
disk is
viscously stable. The thermal instability of this branch can be only validly 
obtained with
the considerations of both the temperature and surface density changes. 
However, our analyses will not affect the results of numerous papers of 
Narayan and
his collaborators (e.g. Narayan \& Yi 1994; Narayan \& Yi 1995; Narayan 1996) 
because most of them mainly considered the optically thin, advection dominated
 accretion disk which is both thermally and viscously stable. Such kind of disk
has been adopted to explain the quiescent state of black hole X-ray binaries
(Narayan, McClintock \& Yi 1996; Yi et al., 1996).

Although the local thermal instability of an optically thick, advection 
dominated disk is probably related to some fluctuations and flickerings 
observed in black hole candidates and AGN (see KAC), we should mentioned that 
the 
global and non-linear evolution properties of a 
thermally unstable
 disk may be different from those in our analyses. We have restricted our 
 analyses in the local
  approximation throughout this work. However,
we think that the surface density change needs to be also seriously considered
in investigating the properties of thermally unstable disks. We expect the 
future work will 
give us
a more clearly understanding to the correlations among the global and 
time-dependent changes
 of the
surface density and other disk parameters.

\acknowledgments
I thank the anonymous referee for many helpful suggestions and comments, and 
Yongheng Zhao for valuable discussions. This work is partially 
supported by the 
Postdoc Science Foundation of China.

\thebibliography{}

\item{} Abramowicz, M.A., Chen, X., Kato, S., Lasota, J.-P., Ragev, O. 1995, 
ApJ, 438, L37
\item{} Chen, X., Abramowicz, M.A., Lasota, J.-P., Narayan, R., Yi, I. 1995, 
ApJ, 443, L61
\item{} Kato, S., Abramowicz, M.A., Chen, X. 1996, PASJ, 48, 67 (KAC)
\item{} Lightman, A., Eardley, D. 1974, ApJ, 187, L1
\item{} Narayan, R. 1996, ApJ, 462, 136
\item{} Narayan, R., McClintock, J.E., Yi, I. 1996, ApJ, 457, 821
\item{} Narayan, R., Yi, I. 1994, ApJ, 428, L13
\item{} Narayan, R., Yi, I. 1995, ApJ, 444, 231
\item{} Novikov, I.D., Throne, K. 1973, in ``Black Holes'', ed. DeWitt, C., 
DeWitt, B. (New York: Gordon and Breach), 343
\item{} Piran, T. 1978, ApJ, 221, 652
\item{} Pringle J.E., 1976, MNRAS, 177, 65
\item{} Pringle, J.E., Rees, M.J., Pacholczyk, A.G. 1973, A\&A, 29, 179
\item{} Shakura, N.I., Sunyaev, R.A. 1973, A\&A, 24, 337
\item{} ----------. 1976, MNRAS, 175, 613
\item{} Shapiro, S.L., Lightman, A.P., Eardley, D.N. 1976, ApJ, 204, 187
\item{} Wu, X.B., Li, Q.B., 1996, \apj, 469, 679 (Paper I)
\item{} Yi, I., Narayan, R., Barret, D., McClintock, J.E., 1996, A\&AS, 120, 
187
\newpage
\begin{center}
{ \bf FIGURE CAPTIONS}
\end{center}
\figcaption{The stability properties of an optically thin disk are shown in 
(a) and the 
relative changes of surface density associated with the unstable thermal mode
are shown in (b). Here we take $\alpha=0.01$, $\Omega=\Omega_k$, $H/r=0.01$
and the Mach number $m=0.01$. The solid, dashed and dotted lines
in (a) correspond to the thermal mode, viscous mode and acoustic modes and
those in (b) represent $\delta_{\Sigma T}$, $\delta_{\Sigma W}$ and 
$\delta_{\Sigma H}$ respectively.} 
\figcaption{The stability properties of an optically thick, radiative cooling 
and radiation 
pressure dominated disk and the relative changes of surface density associated 
with the unstable 
thermal mode. The lines have the same meanings and the parameters are taken 
the same values as 
in Fig. 1.} 
\figcaption{The stability properties of an optically thick, advection 
dominated disk  and the 
relative changes of surface density associated with the unstable thermal mode 
(Note the quantities 
shown in the vertical axes in both (a) and (b) are different from those in 
Fig. 1 and Fig. 2).
 Here we take  $\alpha=0.001$, $\Omega=\Omega_k$, $H/r=0.6$
and the Mach number $m=0.01$. In (a) the solid and dashed lines correspond to
the thermal and viscous modes in the disk without thermal diffusion and the
dotted 
and dot-dashed 
lines correspond to the thermal and viscous modes in the disk with
thermal diffusion. In (b) the solid, long-dashed and short dashed lines 
represent $\delta_
{\Sigma T}$, $\delta_{\Sigma W}$ and 
$\delta_{\Sigma H}$ in the case without thermal diffusion respectively, and 
the dotted, 
dot-long dashed and 
dot-short dashed lines represent those in the case with thermal diffusion.}
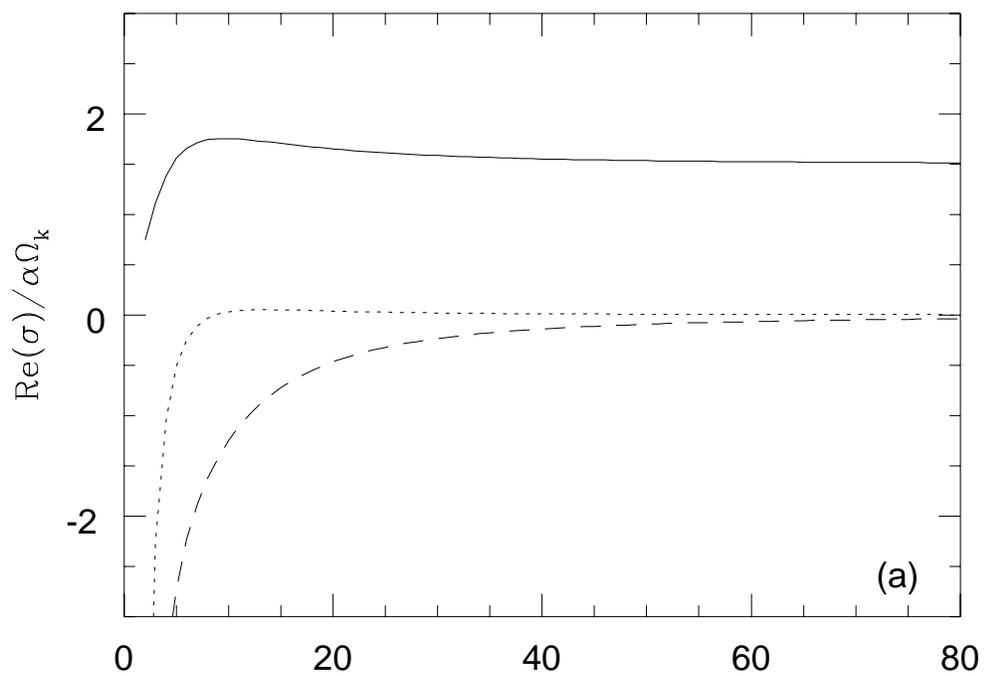
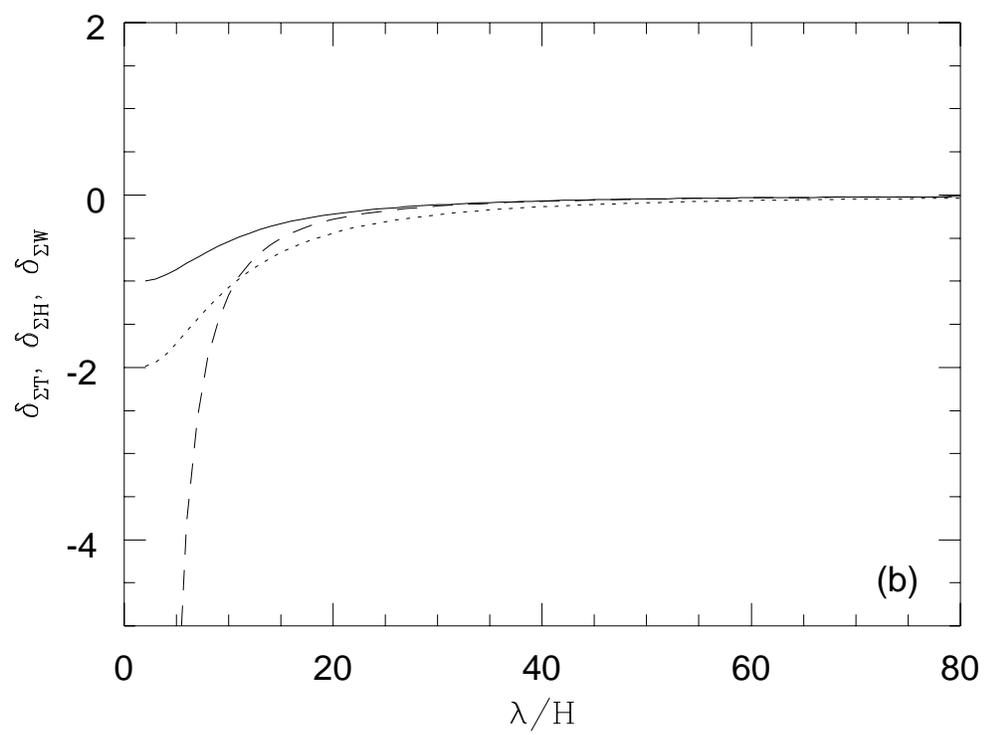
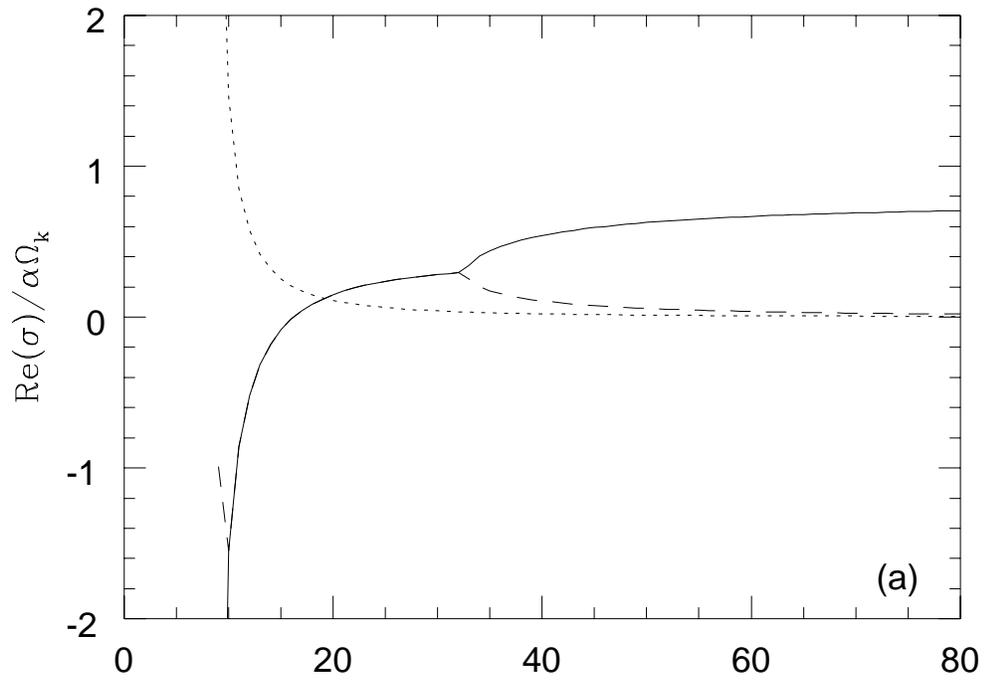
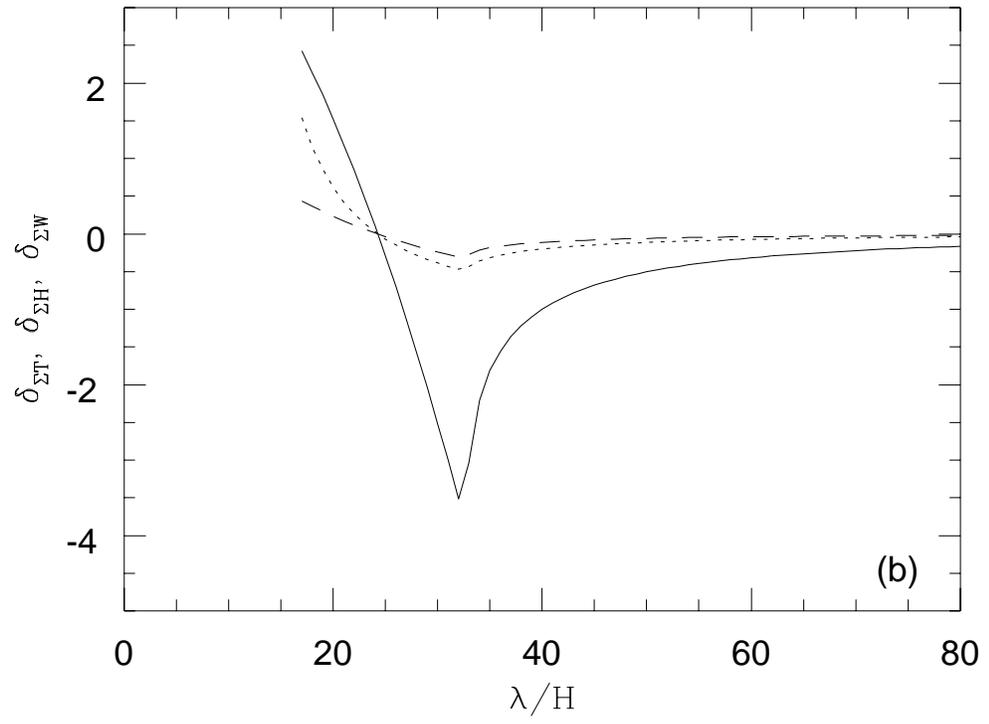
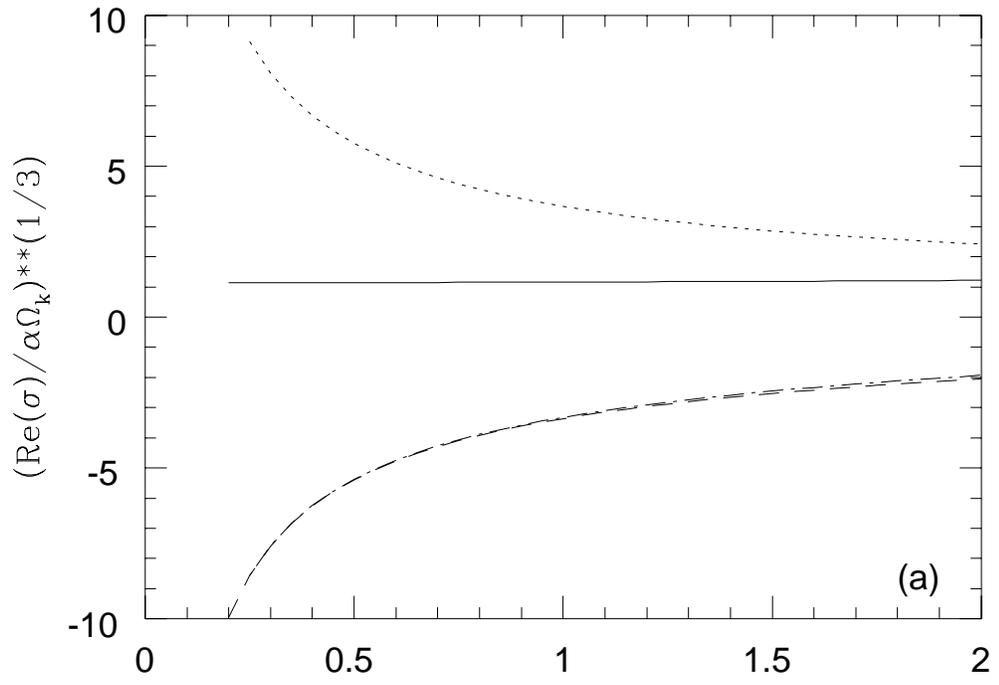
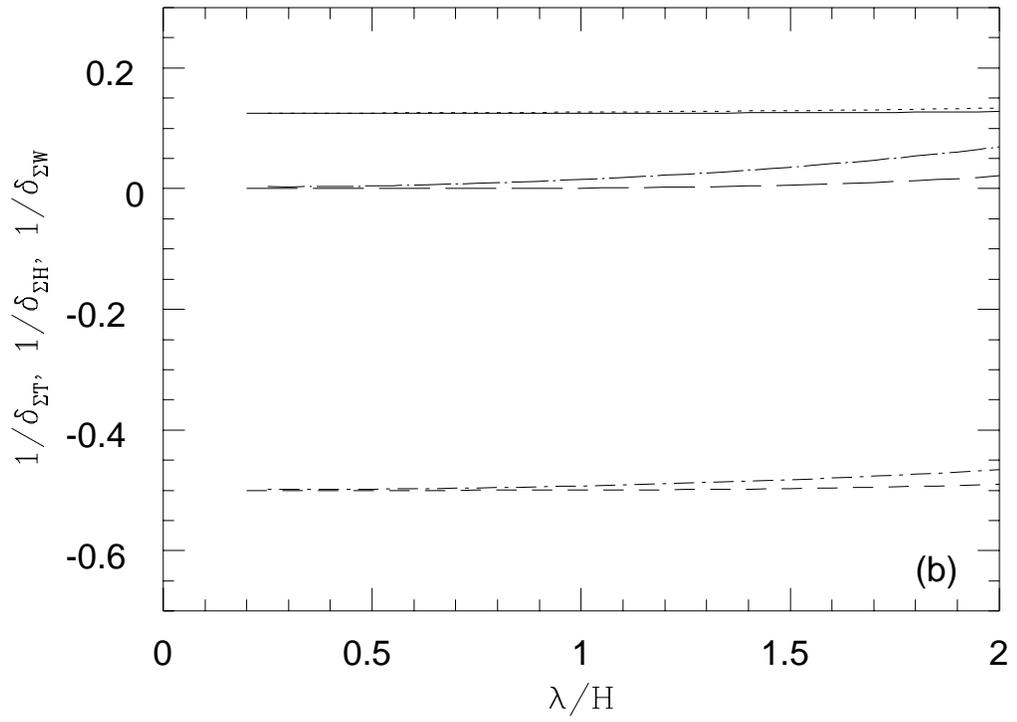
\end{document}